\documentclass[twoside]{article}

\catcode`\@=11 \long\def\@makefntext#1{
\protect\noindent \hbox to 3.2pt {\hskip-.9pt
$^{{\eightrm\@thefnmark}}$\hfil}#1\hfill}       

\def\thefootnote{\fnsymbol{footnote}}
\def\@makefnmark{\hbox to
0pt{$^{\@thefnmark}$\hss}}

\def\ps@myheadings{\let\@mkboth\@gobbletwo
\def\@oddhead{\hbox{}
\rightmark\hfil\eightrm\thepage}
\def\@oddfoot{}\def\@evenhead{\eightrm\thepage\hfil
\leftmark\hbox{}}\def\@evenfoot{}
\def\sectionmark##1{}\def\subsectionmark##1{}}

\def\qed{\hbox{${\vcenter{\vbox{ 
height
0.4pt\hbox{\vrule width 0.4pt height 6pt
\kern5pt\vrule width
0.4pt}\hrule
height 0.4pt}}}$}}

\renewcommand{\thefootnote}{\fnsymbol{footnote}} 

\input{mpla1.sty}

\def\@refcitex[#1]#2{\if@filesw\immediate\write\@auxout
    {\string\citation{#2}}\fi
\def\@citea{}\@refcite{\@for\@citeb:=#2\do
    {\@citea\def\@citea{, }\@ifundefined
    {b@\@citeb}{{\bf ?}\@warning
    {Citation `\@citeb' on page \thepage \space undefined}}
    \hbox{\csname b@\@citeb\endcsname}}}{#1}}
 \def\@refcite#1#2{{[{#1}]\if@tempswa\typeout     
        {WSPC warning: optional citation argument
    ignored: `#2'} \fi}}
 \def\refcite{\@ifnextchar[{\@tempswatrue
    \@refcitex}{\@tempswafalse\@refcitex[]}}

\usepackage{latexsym}
\usepackage{amssymb}
\def\barr{\begin{array}}
\def\earr{\end{array}}
\def\berr{\begin{eqnarray}}
\def\err{\end{eqnarray}}
\def\berrno{\begin{eqnarray*}}
\def\errno{\end{eqnarray*}}
\def\be{\begin{equation}}
\def\ee{\end{equation}}
\def\fr{\frac}
\def\la{\langle}
\def\ra{\rangle}
\def\var{\varphi}

\newcommand{\dder}[2]{\frac{\partial{#1}}{\partial{#2}}}
\newcommand{\dd}[1]{\frac{\partial}{\partial {#1}}}
\newcommand{\tens}[1]{\buildrel {\textstyle \rightrightarrows}\over {#1}}
\newcommand{\pol}[1]{\stackrel{\rm LCP}{\mathrm{RCP}}}

\newcommand{\half}{\fr{1}{2}}
\newcommand{\quart}{\fr{1}{4}}
\newcommand{\no}{\nonumber}

\renewcommand{\a}{\alpha}
\renewcommand{\b}{\beta}

\renewcommand{\d}{\delta}

\renewcommand{\l}{\lambda}

\renewcommand{\r}{\rho}

\renewcommand{\t}{\theta}
\renewcommand{\v}{\varphi}
\renewcommand{\k}{\kappa}
\newcommand{\p}{\psi}
\newcommand{\del}{\partial}

\newcommand{\Schw}{Schwartzchild\,\,}
\newcommand{\La}{Lagrangian\,\,}
\newcommand{\Tul}{Tulczijew\,\,}
\newcommand{\Pap}{Papapetrou\,\,}
\newsavebox{\savepar}

\begin{document}
\setlength{\textheight}{7.7truein} 

\runninghead{Motion of a vector particle$\ldots$}
{Motion of a vector
particle
$\ldots$}

\normalsize\textlineskip \thispagestyle{empty}
\setcounter{page}{1}

\copyrightheading{} 

\vspace*{0.88truein}

\fpage{1} \centerline{\bf MOTION OF A VECTOR PARTICLE
IN A CURVED SPACETIME. II}
\baselineskip=13pt \centerline{\bf FIRST ORDER CORRECTION
TO A GEODESIC IN A SCHWARTZCHILD BACKGROUND.}
\vspace*{0.37truein} \centerline{\footnotesize ZAFAR
TURAKULOV\footnote{\mbox{E-mail:zunn@tps.uz}}}
\baselineskip=12pt
\centerline{\footnotesize\it National University of Uzbekistan, Tashkent,}
\centerline{\footnotesize\it Institute of Nuclear Physics,}
\baselineskip=10pt
\centerline{\footnotesize\it Ulugbek, Tashkent
702132, Uzbekistan and}
\centerline{\footnotesize\it Institute for Studies in
Theoretical Physics and Mathematics,}\centerline{\footnotesize\it
School of Physics, Tehran, Iran.}
\vspace*{10pt}

\centerline{\footnotesize MARGARITA
SAFONOVA\footnote{E-mail:
rita@ipm.ir}} \baselineskip=12pt
\centerline{\footnotesize\it  Institute for Studies in
Theoretical Physics and Mathematics,} \baselineskip=10pt
\centerline{\footnotesize\it
School of Physics,
P.~O.~Box 19395-5531 Tehran, Iran.} \vspace*{0.225truein}

\publisher{(received date)}{(revised date)}
\vspace*{0.21truein}
\abstracts{
The influence of spin on a photon's motion in a \Schw and FRW spacetimes
is studied. The first order correction to the geodesic motion is found. 
It is shown that unlike the world-lines of spinless particles, the photons
world-lines do not lie in a plane.}{}{}




\setcounter{footnote}{0}
\renewcommand{\thefootnote}{\alph{footnote}}

\vspace*{1pt}\textlineskip

\section{Introduction}
The motion of spinning particles, test ones or tops, in a curved
spacetime has been a subject of great interest for the last
nearly 80 years. It is amuzing to note that, even today, there is
no general consensus on the behaviour of particles with spin in
external gravitational fields. It seems that the first to discuss
the motion of free extended material particles with spin in an
external electromagnetic field was Frenkel\cite{frenkel}. The
classical test body was considered to be so small compared with
the background curvature length-scale that all its multipoles
beyond the dipole could be neglected. The equations of the free
motion of such a `pole-dipole' particle for the linearized theory
of gravitation were obtained by Mathisson\cite{mathisson}.
Papapetrou\cite{pap} worked with non-singular localized
energy-momentum tensor $T_{ij}$ for a spinning particle (in the
limit of a vanishing mass) and obtained equations for moments of
$T_{ij}$. Dixon\cite{dixon} found a way to make Papapetrou's type
of argument covariant at each step of the derivation and also
found equations which referred to paths described by an arbitrary
parameter rather than the world-line length. He came up with the
equations of motion in the form 
$$
\fr{Dp^i}{Dq}=\fr{1}{2}R^i_{\,\,jkm}v^jS^{km}\,,$$
\be
\fr{DS^{ij}}{Dq}=p^{[i}v^{j]}\,,
\label{eq:dixon}
\ee
with $q$ an arbitrary
parameter along the curve, $v^i=dx^i(q)/dq$, $D/Dq$ a covariant
derivative along the world-line $x^i(q)$ and $p^i$ and $S^{ij}$
defined as in Papapetrou\cite{pap}. The explicit form of generalized 
momentum $p$, containing the additional term depending
only on spin and connection, was shown in \refcite{turakulov}, 
in complete agreement with the definition of Papapetrou.   
In the derivation of these equations by Dixon nothing required 
the object to have a rest mass or required $g_{ij}dx^idx^j$ to be nonzero. 
They were, therefore, sufficiently general to relate to any localized object.

However there have not yet been satisfactory direct derivation of
these equations for massless particles with spin. In \refcite{baylin}
Baylin and Ragusa treated massless
particles as traceless, but the authors couldn't come out with the
final answer whether "traceless"-massless particles' motion
deviate from null geodesics. At the same time they mention that
"we have been unable to prove that null geodesics necessarily
follow for the general case of spin in curved spacetime. It even
seems unlikely in view of complexity of [Papapetrou-Dixon]
equations". 

Explicit coupling to curvature actually turns out to
be the general rule in the field theoretical case, so that
strictly geodesic behaviour is not to ever be expected in general.
Because fields are defined at all spacetime points and not just
along the trajectories associated with point particles, the
associated second order field equations are sensitive to the
values of the connections at differing points. Both Dirac\cite{mannheim}
and Maxwell\cite{eddington,prasanna,mannheim} equations lead to an 
implicit expressly non-inertial
coupling to the Riemann curvature tensor. And even though it is
possible to remove the Christoffel symbols at any given point (or
even along a single curve), it is nonetheless impossible to remove
them from an entire region, with the curvature dependent terms we
find in the second order wave equations being the field theoretic
generalization of the (covariantly describable) geodesic deviation
found for pairs of nearby falling particles.

In the previous paper (paper~I\cite{paperI}) we have derived
the equations of motion for a vector particle with spin from a
Lagrangian. We have worked out the approach that gives a
satisfactory approximation to the wave as some curves that can be
called "rays" and, at the same time, takes into account the
polarization. The obtained equation coincided with the Papapetrou
equation and is independent of a mass of a particle. In the
present work we apply equations derived in paperI for photons
transversely propagating in a \Schw metric. We develope the
technique to calculate the corrections to the geodesic motion. We
show the deviation from a geodesic in \Schw metric and examine 
the consequences of the deviation in different gravitational backgrounds. 
We present a general discussion of the results.

\section{Geodesic congruences for \Schw metric}
\subsection{Hamilton-Jacobi Equation for photons}

The Hamilton-Jacobi equation (HJE) for photons reduces to the
eikonal equation,
\be
g^{ij} \frac{\del \psi}{\del x^i}\fr{\del \p}{\del x^j}=0\,\,.
\label{eq:eikonal}
\ee
In \Schw \/ geometry
$$
ds^2=\left(1-\fr{2m}{r}\right)\,dt^2 - \left(1-\fr{2m}{r}\right)^{-1}
\,dr^2 - r^2 \,d\theta^2-r^2\sin^2{\theta} \,d\v^2\,\,,
$$
with the coordinates: $x^{\mu}=\{t,r,\theta,\v\} = \{0,1,2,3\}$,
the metric coefficients are:
\be
g^{ij} = {\rm diag}\left(\frac{1}{1+2\phi},- (1+2\phi),-\fr{1}{r^2},
-\fr{1}{r^2\sin^{2}{\theta}} \right)\,\,,
\label{eq:coeff}
\ee
where by $\phi$ we denoted the ``Newtonian" potential $\phi=-m/r$. 
We employ here the Planck units $c=\hbar=G=1$. Eq.~\ref{eq:eikonal} becomes

\be \fr{1}{1+2\phi}\left(\dder{\p}{t}\right)^2 -
(1+2\phi)\left(\dder{\p} {r}\right)^2
-\fr{1}{r^2}\left(\dder{\p}{\t}\right)^2 -
\fr{1}{r^2\sin^2{\theta}}\left(\dder{\p}{\varphi}\right)^2 = 0\,\,.
\label{eq:expanded}
\ee
Following the usual approach to separating the variables in HJE, we
represent the desired solution in the form $$ \p =Et +L_z \varphi +R(r)
+\Theta(\theta)\,\,, $$ with the convention that $\sqrt{a^2}=\pm a$.
Substituting it in Eq.~\ref{eq:expanded}, we obtain
the equations where variables have got separated indeed and produced
an arbitrary constant (the constant of separation, total angular momentum)
$L$:
$$
\fr{E^2 r^2}{1+2\phi} - (1+2\phi) r^2 \left( R' \right)^2 =
(\Theta')^2 + \fr{L_z^2}{\sin^2{\t}} = L^2\,\,,
$$
from where we find functions $\Theta$ and $R$, and
get the solution in the form $$ \p_{\pm} = Et +L_z \varphi
\pm\int \fr{\sqrt{E^2 r^2-L^2 (1+2\phi)}}{r(1+2\phi)}dr\pm\int\sqrt{L^2
-\fr{L_z^2}{\sin^2{\t}}} d \t\,\,. $$

\subsection{Associated Tetrad Basis}

Any coordinate system $\left\{x^i\right\}$ in the spacetime specifies the
natural vector $\left\{\del_i\right\}$ and covector $\left\{dx^i\right\}$
frames at each point. We defined the metric by the scalar products of
natural covectors
$$
g^{ij} = \la dx^i, dx^j \ra\,\,,
$$
because our main requirement to the metric is that it shall separate the HJE for
geodesics, which looks much more simple in the contravariant components
than in covariant ones.

We choose for simplicity the equatorial plane, $\theta = \pi/2$ ($L=L_z=p_{\v}$). 
Consequently, the dynamic phase $\p$ becomes
\be
\p_{\pm}  = E t \mp \int \fr{\sqrt{E^2 r^2 - L^2 (1+2\phi)}}{r (1+2\phi)} dr
\pm  L  \v \,\,,
\label{eq:sol}
\ee
where we fixed the signs as
\berrno
&\p_{+}  = E t - R +  L \v \,\,,\\
&\p_{-}  = E t + R -  L \v \,\,,
\errno
and write the gradients of the phase  
$$
d \p_{\pm} = E dt \pm L d \v \mp R'dr\,.
$$
Now we can construct the associated basis on which the connection form
is identically zero. Taking the scalar product $\la d \p_{+},
d \p_{-}\ra$, we obtain
$$
\la d \p_{+},d \p_{-}\ra = E^2\la dt,dt\ra - L^2\la d\v,d\v \ra -
\left(R'\right)^2 \la dr,dr\ra = \fr{2 E^2}{1+2\phi}\,,
$$
since from the metric coefficients (Eq.~\ref{eq:coeff}) we have $\la dt,dt \ra=
g^{tt}= 1/(1+2\phi)$\,\,,
$\la d\t,d\t \ra =g^{\t\t}=-r^{-2}$, $\la dr,dr \ra= g^{rr} =-(1+2\phi)$ and~
$\la d\v,d\v\ra=g^{\v\v}=-1/r^2\sin^2{\t}$. The first pair of basis vectors is thus
$$ \nu^{\pm} = \fr{d \p_{\pm}}{\sqrt{\la d \p_{+},d
\p_{-}\ra}} = \fr{d \p_{\pm} \sqrt{1+2\phi}}{E \sqrt{2}}\,\,.
$$
Since $\nu^{\pm}$ is a null vector, we need to add timelike vectors to our basis. 
Applying the Hamilton-Jacobi theorem\cite{pars} to the solution (Eq.~\ref{eq:sol}) we
can construct the integral of motion in the form of a
function whose values are constant on the trajectory
$$
V =\fr{\del \p_{\pm}}{\del L} = \pm \left( \v - \fr{\del R}{\del L}
\right)\,\,. $$ We choose $$ d V = d \v - \fr{\del R'}{\del L} dr\,.
$$
The scalar product $\la d V,d V\ra$ is
$$ \la d V,d V\ra  =  \la d\v,d\v \ra + \left(\fr{\del R'}{\del L}
\right)^2 \la dr,dr \ra =  -  \fr{1}{r^2 - D^2 (1+2 \phi)} \,\,,
$$
where we denoted $L/E=D$, with $D$ having the meaning of the impact parameter. Thus,
$$
\nu^2 = \fr{d V}{\sqrt{-\la d
V,d V\ra}}= d V \sqrt{r^2 - D^2(1+2 \phi)}\,\,.
$$
As a fourth basis covector we choose $d \t$:
$$
\nu^3= \fr{d\t}{\sqrt{-\la d \t,d \t \ra}} = r d\t\,\,.
$$
Thus, we constructed the associated basis of 1-forms,
\begin{eqnarray}
&&\nu^{\pm} =
\fr{\sqrt{1+2\phi}}{E\sqrt{2}}\left[Edt\pm Ld\v \mp R'dr\right]\,\no\\
&&\nu^2 = \sqrt{r^2 - D^2(1+2 \phi)}\left[d\v - \dder{R'}{L}dr \right]\,,\label{eq:nu-s}
\\
&&\nu^3 = r d\t\,\,.\no
\err
This basis is orthonormal in the sense of $\la \nu^{\pm},\nu^{\pm} \ra =0$ and  
$\la \nu^{\pm},\nu^{\mp} \ra = 1$. The basis dual to $\left\{\nu^a \right\}$ 
is the vector basis $\left\{\vec n_a \right\}$ ($\nu^a(\vec n_b)=\d^a_b$)
\berr
&& \vec n_{\pm} = \frac{\sqrt{1+2\phi}}{E\sqrt{2}} \left[
\fr{E}{(1+2\phi)} \dd{t} \mp R' (1+2 \phi) \dd{r} \pm \fr{L}{r^2}
\dd{\v} \right], \no\\
&& \vec{n}_2 = \fr{D(1+2\phi)}{r} \dd{r} +
\fr{\sqrt{r^2- D^2(1+2\phi)}}{r^2}\,\dd{\v}\,\,, \\
&& \vec{n}_3 = \fr{1}{r} \, \dd{\t}\,\,.\no
\err
We also introduce orthonormal sub-frames, constructed from 
the associated basis,  
$$
\nu^{0} = \fr{\nu^{+} + \nu^{-}}{2}\,,\,\,\,\,\,
\nu^1 = \fr{\nu^{+} - \nu^{-}}{2}\,;
$$
and 
\berrno
\left\{ \barr{l}
\vec{n}_0 = \fr{1}{\sqrt{2(1+2\phi)}} \, \dd{t}\,,  \\
\\
\vec{n}_1 = \fr{\sqrt{1+2\phi}}{E\sqrt{2}} \left[\fr{L}{r^2}
\,\dd{\v} - (1+2\phi) R' \, \dd{r} \right]\,. \earr
\right.
\errno
According to the definition {\bf Def.~1}
\fbox{\rule[-0.2cm]{0.0cm}{0.7cm}$\left \{ n_b \right \}
\,:\,\,n_b=h^i_b \,\dd{x^i}$} the non-zero tetrad functions of the vector
basis are
$$ \begin{array}{ll}
h_0^t = \fr{1}{\sqrt{2 (1+2\phi)}}  &  \hskip 3cm h_1^r = 
-\fr{\sqrt{1+2\phi}}{r\sqrt{2}} \sqrt{r^2-D^2(1+2\phi)}  \\
 & \\
 h_2^r = \fr{D(1+2\phi)}{r} &  \hskip 3cm h_1^{\var} = \fr{D}{r^2}
\sqrt{\fr{1+2\phi}{2}} \\
& \\
h_2^{\var} = \fr{\sqrt{r^2 -D^2 (1+2\phi)}}{r^2} &  \hskip 3cm h_3^{\t} = \fr{1}{r}  
\end{array} $$

\subsection{Polarization basis on the geodesic}

We assume a narrow beam of rays and neglect transversal
derivatives---in the zeroth approximation polarization basis is
propagating parallely along the ray. We can write tensors as
$\tens{F}=F^{ab}\,f_{ab}$, where $\left\{f_{ab} \right\}$---tensor
basis  
\be
f_{ab}=\vec{n}_{[a} \otimes \vec{n}_{b]}=
\fr{1}{2} \left(h^i_{a} h^j_{b} -
h^i_{b} h^j_{a} \right) \left[ \dd{x^i} \otimes \dd{x^j}
\right]\,, \label{eq:example}
\ee
and square brackets mean antisymmetrization. We define vertical
($v$) and horizontal ($h$) linearly polarized waves  \berr
&&\vec A_v=A^2\,\vec n_2=(a\sin{\p})\, \vec n_2\,\,,\no\\
&&\vec A_h=A^3\,\vec n_3=(a\cos{\p})\,\vec n_3 \,, \label{eq:forms}
\err
where $a$ is an amplitude. The waves with left and right circular
polarizations (LCP and RCP, correspondingly) are their linear combinations,
\berr
&&\vec A_{\rm R} =\vec A_v+\vec A_h=(a\sin{\p})\,\vec n_2+(a\cos{\p})\,\vec n_3 \,\,,\no\\
&&\vec A_{\rm L} = \vec A_v-\vec A_h=(a\sin{\p})\,\vec n_2-(a\cos{\p})\,\vec n_3\,.
\label{eq:alphas}
\err
The derivatives of (\ref{eq:forms}) are
\berr
&& \dot{\vec A_v}=\dot A^2 \vec n_2 =(a \dot{\p}\cos{\p})\,\vec n_2=
\left(\fr{aE\sqrt{2}}{\sqrt{1+2\pi}}\cos{\p}\right)\,\vec n_2 \,,\no\\
&&\dot{\vec A_h}=\dot A^3\,\vec n_3=
-(a \dot{\p}\sin{\p})\,\vec n_3=
-\left(\fr{aE\sqrt{2}}{\sqrt{1+2\pi}}\sin{\p}\right)\,\vec n_3 \,,
\label{eq:derivatives}
\err
where we used the fact that on the geodesic the connections are zero and
basis vectors $\vec n_a$ are parallely transported (by construction). The operator
'{\bf dot}' was defined in the following way. By the definition, a gradient
of a function in dual frames is
$$
df\equiv \left(\vec n_a \circ f\right)\nu^a\,.
$$
We take the derivative only along the beam, hence,
$$
df=\left(\vec n_{-} \circ f\right)\nu^+\,,
$$
and replacing $\nu^{+}$ from Eqs.~\ref{eq:nu-s},
$$
df=\left(\vec n_{-} \circ f\right)\fr{\sqrt{1+2\phi}}{E\sqrt{2}}d\p^+\,.
$$
Finally, the 'dot' operator takes the form
$$
\dot f\equiv \fr{df}{d\p^+}=\left(\vec n_{-} \circ f\right)\fr{\sqrt{1+2\phi}}{E\sqrt{2}}\,.
$$
The spin tensor is defined as (see Appendix A)
\be
S_{ab} =\half \int
\left[\dot{A}_2 A_3-\dot{A_3} A_2\right]d^3x \,\,.
\label{eq:spin_definition}
\ee
The only non-zero component of the spin current 
$$
J^0_{23}=\fr{a^2E}{\sqrt{2(1+2\phi)}}\,,
$$
and the only non-zero component of spin for the LCP
polarization we find from
Eqs.~\ref{eq:alphas},~\ref{eq:derivatives}~and~\ref{eq:spin_definition}
$$
^{\rm L}S_{23}=\int \fr{a^2E}{\sqrt{2(1+2\phi)}} \left(\cos^2{\p} +
\sin^2{\p}\right)\, d^3x= \int \fr{a^2E}{\sqrt{2(1+2\phi)}}\, d^3x
=\mbox{const}=1\,.$$
For RCP, correspondingly, the sign is opposite,
$$ ^{\rm R}S_{23}=-1\,. $$

It is convenient to introduce the following notation for the spin-gravity
coupling term, the tensor
$$ H^i_{\,\,j} = \fr{1}{2}R^i_{\,\,jkl}S^{kl} \,\,, $$
where in order to express the spin tensor in the coordinate frame, we use the
tensor basis defined in Eq.~\ref{eq:example}, 
$$
S_{23}=\pm 1=S^{23}=S^{kl}h^{[2}_k h_l^{3]}\equiv \pm\left(f_{kl}\right)^{23}\,,
$$
and
$$
S^{kl}=S^{23}\left(f_{23}\right)^{kl}\,.
$$
From six components of $\{f_{23}\}$ basis only two survive
\berrno
&&\left(f_{23}\right)^{\t r} = -\fr{1}{2}h^{\t}_3 h^{r}_2 =
-\fr{D(1+2\phi)}{2r^2}\,,\\
&& \\
&& \left(f_{23}\right)^{\t\v} = -\fr{1}{2}h^{\t}_3 h^{\v}_2 =
-\fr{\sqrt{r^2-D^2(1+2\phi)}}{2r^3}\,,
\errno
where the tetrad functions are found from {\bf Def.~1}).

Thus, for LCP the components of this tensor are
\berrno
&&^{\rm L}H^{\t}_{\,\,r}=\half R^{\t}_{\,\, r\t r} \left(f_{23}\right)^{\t r} =
-\fr{mD}{4r^5}\,\,;\no\hskip 8cm \\
&&\\
&&^{\rm L}H^{\t}_{\,\,\v}=\half R^{\t}_{\,\, \v \t \v} \left(f_{23}\right)^{\t \v}=
-\fr{m\sqrt{r^2-D^2(1+2\phi)}}{2r^4} \,\,,
\errno
where the corresponding components of the Riemann tensor are
$$
R^{\t}_{\,\, r\t r}=-\fr{m}{r^3(1+2\phi)},
$$
$$
R^{\t}_{\,\, \v \t \v}\fr{2m}{r}\,.
$$

\section{The Papapetrou Equation}

The first equation of (Eq.~\ref{eq:dixon}) can be written in the form\be
\dot p^i=\fr{1}{2}R^i_{\,\,jkl}S^{kl}\dot x^j\,.
\label{eq:papapetrou}\ee With our definition for the circular polarization
the equation transforms to\be
\dot p^i=\pm H^i_{\,\,j}\dot x^j(S^{23})
\label{eq:variation}
\ee
with `+' corresponding to LCP and `-' -- to RCP. The geodesics we are
considering lie in the equatorial orbit $\t=\pi/2$ and, hence,
$\dot \t=0$. Since the only surviving $H$ tensors contain $\t$ components,
in the Eq.~\ref{eq:variation} we have non-zero right hand side only for
$\t$-component of the momentum. We are considering the deviation $\d x^i(\p)$
from the geodesic
$x^i(\p)$ as our first approximation, $\d x^i(\p)\equiv(0,0,\d\t(\p),0)$. The
Papapetrou equation for all coordinates but $\t$ coincides with the
geodesic equation, and for the $\t$ coordinate we have:
$$
\dot p^{\t}=0\,,
$$
$$
\d\dot p^{\t}  = \pm H^{\t}_j\dot x^j\,,
$$
where in the zeroth approximation (geodesic motion) the spin-dependent term of
the generalized momentum $p^i$ vanishes. To find the components of the
velocities we need to find the geodesics in the explicit parametrized form. If
we express the solution as $\p(x^i,\a_a)$, where $x^i$ represents
variables and $\a_a$ represents constants, then equations
$$
\dder{\p}{\a_a} = \p_a(x^i,\a_a)=\mbox{const}\,\,.
$$
give the analytical representation of geodesics in parametrized form
(Hamilton-Jacobi theorem\cite{pars}). Applying this to the Eq.~\ref{eq:sol}
gives geodesics in the form of the following system of equations
with $r$ as an independent parameter:
\berr
\dder{\p_{\pm}}{E}=t_0\,,\no \\
&t-t_0=\pm \int \fr{Erdr}{(1+2\phi)\sqrt{E^2r^2-L^2(1+2\phi)}} \equiv t(r)\,;\no \\
& \label{eq:parametric}\\
\dder{\p_{\pm}}{L}=\v_0\,,\no \\
&\v - \v_0 =\pm \int \fr{Ldr}{r\sqrt{E^2r^2-L^2(1+2\phi)}} \equiv \v(r)\,. \no
\err
Now we need to find the relation between momentum and velocities. The components 
of the momentum vector are obtained from the solution of HJE,
$$
p_t=\dder{\p}{t}=E\,;\,\,\,p_r=R^{\prime}\,;\,\,\,p_{\v}=L\,;
\,\,\,p_{\t}=0\,,
$$
and, accordingly,
$$
p^t=\fr{E}{1+2\v}\,;\,\,\,p^r=-(1+2\v)R^{\prime}\,;
\,\,\,p^{\v}=-\fr{L}{r^2}\,;\,\,\,p^{\t}=0\,.
$$
Now we find $(x^i)^{\prime}$-s from (\ref{eq:parametric})
\berr
&&(x^t)^{\prime}=\fr{Er}{(1+2\phi)\sqrt{E^2r^2-L^2(1+2\phi)}}\,\,;\,\,\,\,
(x^r)^{\prime}=1\,\,;\,\,\,\, (x^{\t})^{\prime}=0\,;\no\\
&&(x^{\v})^{\prime}=\fr{L}{r\sqrt{E^2r^2-L^2(1+2\v)}}\,,
\label{eq:primes} \err where `prime' denotes differentiation
with respect to (w.r.t) $r$. By taking out the common multiplier, we
express the momentum through velocities $$
p^i=\fr{\sqrt{E^2r^2-L^2(1+2\phi)}}{r}\,\left(x^i\right)^{\prime}=
\fr{E\sqrt{r^2-D^2(1+2\phi)}}{r}\,\left(x^i\right)^{\prime}\,. 
$$
In our first approximation we can assume the same relation between
the velocity and the momentum as on the geodesic, 
\be 
\d p^{\t}=
\fr{E\sqrt{r^2-D^2(1+2\phi)}}{r}\,\d\t^{\prime}\,. \label{eq:p}
\ee 
By `prime' we mean the covariant differentiation with respect to
$r$. However, we can express $\d p^{\t}$ through our associated
basis, 
\be 
\d p^{\t}= \sqrt{-g^{\t\t}}\d p^3=\fr{1}{r}\d p^3\,.
\label{delta_p^3} 
\ee 
According to the zero-connection property of
the associated basis, we can replace the covariant differentiation
of $\d p^3$ w.r.t.~$r$ with ordinary differentiation w.r.t.~$r$.
The same would hold for a product of $\d p^3$ with any scalar
function, as in (\ref{delta_p^3}). Hence, we write the Papapetrou
equation as $$ \left(\d p^{\t}\right)^{\prime}=H^{\t}_{\,\,j}
\left(x^j\right)^{\prime}\,.$$ 
The contraction with the velocity gives the right-hand side of the equation 
$$
H^{\t}_{\,\,j}
\left(x^j\right)^{\prime}= H^{\t}_{\,\,r}
\left(x^r\right)^{\prime}+ H^{\t}_{\,\,\v}
\left(x^{\v}\right)^{\prime}=\fr{mD}{4r^5}\,, 
$$
with $(x^j)^{\prime}$-s from Eq.~\ref{eq:primes}. Thus, 
the Papapetrou equation is 
\be 
\left(\d p^{\t} \right)^{\prime}=\fr{mD}{4r^5}\,, \ee and, according to
(\ref{eq:p}), 
$$ 
\left( \fr{E\sqrt{r^2-D^2(1+2\phi)}}{r}\d\t^{\prime} \right)^{\prime}
= \fr{mD}{4r^5}\,. 
$$ 
Thus, the
trajectory of a photon world-line is: $t(r)$, $\v(r)$, $\t (r)=
\pi/2 + \d \t(r)$, where for LCP (+) and RCP (-), correspondingly,
\be 
\pm\d \t (r)= -\fr{mD}{16E} \int^r
\fr{dr}{r^3\sqrt{r^2-D^2(1+2\phi)}}\,, 
\label{eq:phi-integral} 
\ee
with gravitational deflection term $\v (r)$ defined in
(\ref{eq:parametric}). With $(1+2\phi) \equiv (1-r_g/r)$ and $m
\equiv 2r_g$, Eq.~\ref{eq:phi-integral} takes the form 
\be 
\pm \d\t (r)=-\fr{r_g D}{8E} \int \fr{dr}{r^3\sqrt{r^2-D^2(1-r_g/r)}}\,.
\label{eq:delta_phi} 
\ee 
This formula shows the effect of dispersion, the energy-dependent
deviation from a geodesic motion due to the spin. 
To see whether this quantity will have any appreciable effect on
the propagation of light, we have to make an estimate of the order of
magnitude of this integral. It is more convenient to express the
result in terms of the distance $r_0$ of closest approach, rather
than the impact parameter $D$. At $r_0$
$$
R'=\pm \fr{\sqrt{E^2 r^2 - L^2 (1+2\phi)}}{r (1+2\phi)}=0\,\,,
$$
$$
r^2 - D^2 (1-\fr{r_g}{r})=0\,,
$$
Since the difference between $D$ and $r_0$ is small, we write $D=r_0 + \d r_0$,
and inserting it into the previous equation gives to the
first order $\d r_0 \approx r_g r_0/2(r_0-r_g)$ and $D\approx (2r_0+r_g)/2$.
Using $r_0$ instead of $D$ in the integral (\ref{eq:delta_phi}) gives
\berrno
&&\pm\d\t(r)=-\fr{1}{8E}\int \fr{r_0 r_g dr}
{r^3\sqrt{r^2-\fr{(2r_0+r_g)^2}{4}\left(1-\fr{r_g}{r}\right)}}=\\
&&-\fr{1}{8E}\int \fr{r_0 r_g dr}
{r^3\sqrt{r^2-r_0^2}}\left(1+\fr{r_0r_g}{2r(r^2-r_0^2)}\right)\,.
\errno 
Up to the first order in $r_g/r$ we obtain ($E=1/\l$)
$$
\pm\d\t(r)=-\fr{r_g\l}{16r_0^2}\left[\sin^{-1}\left(\fr{r_o}{r}\right)-
\fr{r_0\sqrt{r^2-r_0^2}}{r^2}\right]\,. $$
And at the observer the deviation from the $\t=\pi/2$ plane is 
$$
^L_R\d\t(+\infty)=\mp\fr{\pi r_g \l}{16r_0^2}\,.
$$
We notice that at $r_g\rightarrow 0$, or at $E\rightarrow\infty$,
the spin-dependent deviation vanishes. Spinless particles always move 
on plane orbits, in the case of a circularly polarized photon with 
the spin (helicity) parallel to the velocity, the perturbed motion is not plane. 

We feel that we shall comment here on the problem of the deflection of
light. Contrary to the reports\cite{vishwa,drummond} about the 
polarization dependence of light deflection in a 
\Schw metric, and in concordance
with the pioneering work of Corinaldesi and Papapetrou\cite{cor-pap}, we conclude
that spin gives no contribution to the deflection (the deflection
(Eqs.~\ref{eq:parametric}) is identical with that which is found from using the
equations of geodesics). 

Of course, the result with $\l/r_0$ has an extremely small value and it is 
evident that the effect certainly has no observational consequences in the
gravitational fields appropriate to present-day astrophysics and cosmology. 
Nevertheless, we feel that the main significance of the result is theoretical 
and does not depend on its observability. Moreover, in view of recent
findings of the appearence of a chaotic behaviour in the motion of a 
spinning particle in a \Schw spacetime due to the spin-orbit 
coupling\cite{chaos}, and of a report of a measurement of a time delay between
RCP and LCP signals from a pulsar PSR 1937+21\cite{thorsett}, we believe 
that a study of spin effects on the orbital evolution of 
relativistic systems is very important from the viewpoint of observations 
as well as from an academic one.
 
\subsection{Robertson-Walker Background and Radial Motion}

From the general form of the Eq.~\ref{eq:papapetrou} it is clear that
in the flat spacetime ($R_{\mu\nu\k\r}=0$) or for the spinless particle
($S^{\mu\nu}=0$), the equation reduces to the geodesic
equation or, in case of a massless particle,
to the null geodesic equation.

In general metric at every point of a spacetime there exists a
Weyl principal tetrad in which the Riemann tensor takes its normal
form\cite{kramer} (it is strictly diagonal in pairs of indices).
If the direction of the momentum of a spinning particle
coincides with one of the principal tetrad directions, then due to the \Tul
constraint\cite{tul} (one of the supplementary conditions to the
equations of motion, introduced in order to reduce the number of
spin independent degrees of freedom, $p_{\a}S^{\a\b}=0$) the
Papapetrou force is identically zero. In the homogeneous
gravitational background, or background with constant curvature,
any direction is the principal direction, and hence the \Pap force
is always zero in such metrics. Since the FRW spacetime is
isotropic and homogeneous, the origin of polar spherical
coordinate system, for example, can be any arbitrary point. If we
place the origin on a geodesic, the geodesic becomes strictly
radial in this coordinate system. Thus, the photon's motion is 
geodesic in FRW spacetime. The same can be shown for any
spacetime of constant curvature, for example, de-Sitter,
the fact first discovered by Bailyn and Ragusa \refcite{ragusa}.

The radial direction in \Schw metric is both principal (in the sense of
Weyl principal tetrad) and geodesic and, thus, there is no deviation. 
The motion is just pure radial geodesic with
\berrno
&& t(r)=t_0 \pm \int\fr{dr}{1+2\phi}\,,\\
&& R(r)= \pm \int\fr{Edr}{1+2\phi}\,,\\
&&\t(r)=\mbox{Const}\,,\no\\
&&\v(r)=\mbox{Const}\,.
\errno

\vspace*{1pt}\textlineskip \section{Acknowledgments}
\vspace*{-0.5pt}
Authors would like to thank the Institute for Studies in Theoretical 
Physics and Mathematics in Tehran, Iran, for financial support as well as  
for its kind hospitality while this work was concluded.

\appendix{ Spin tensor derivation}
In paper~I we have defined the spin tensor as following $$
mS_{ab}=\half\left(\dot{A}_a A_b - \dot{A}_b A_a\right)\,. $$
However, it was pointed to us recently\footnote{We express our
thanks to Dr.~M.~M.~Sheikh-Jabbari from IPM, Tehran, Iran.}\,\, that
in that definition the dimensions are not correct. Removing the
mass term restores the correct dimensionality. It should be noted,
however, 
that neither the derivations of the equations of motion for spin
and momentum nor the final results of the paper~I suffer from this
redefinition. The derivation of the equation of motion of spin
does not depend on the mass term, and in the last equation for the
momentum (Eq.~31, paper~I) the mass term is absorbed in the
momentum term $$ g_{ij}m\fr{D\dot{x}^j}{Ds}\equiv
g_{ij}\fr{Dp^j}{Ds}= \dot{x}^jR^k_{\,\,jil}S^l_{\,\,k}\,. $$ We
will present here the proper derivation of the spin tensor which
was just postulated in paper~I. Using the Noether theorem we will
derive the spin tensor as a current conserved under the rotation
of a local orthonormal frame as an operation of internal symmetry
(see, for example, \refcite{itzykson}). The field \La 
is $$ 2{\cal L}_{\rm F}=-\dot{\bf{A}}^2+m^2\bf{A}^2\,, $$
where `dot' means covariant derivative. Under the infinitesimal
rotation of the frame $\d\a^a_{\,b}$, the new \La is the function
of $$ \tilde{{\cal L}}_{\rm F}=\tilde{{\cal L}}_{\rm
F}(\dot{\bf{A}},\bf{A}, \d\a,\d\dot{\a})\,, $$ where
$$
\d{\bf A}=\d\a^{ab}\hat{S}_{ab}{\bf A}\,\,.
$$
Thus, the new \La is
$$
2\tilde{{\cal L}}_{\rm F}={\cal L}_{\rm F}+\d\left(-\dot{\bf{A}}^2+m^2\bf{A}^2
\right)\,.
$$
Let us compute it,
\berrno
&&\tilde{{\cal L}}_{\rm F}=-\half\left(\bf{A}+\d\bf{A}\right)^{\bullet 2}+
\fr{m^2}{2}\left(\bf{A}+\d\bf{A}\right)^2=\\
&&-\half\left(\dot{\bf{A}}+\left(\d A\right)^{\bullet}\right)
\left(\dot{\bf{A}}+\left(\d\bf{A}\right)^{\bullet}\right) +
\fr{m^2}{2}\left(\bf{A}^2+2\left(\d\bf{A}\right) \bf{A}\right)=\\
&&\\
&&{\cal L}_{\rm F}-\dot{\bf A}\left(\d\a^{ab}\hat{S}_{ab}{\bf
A}\right)^{\bullet}\,\,. 
\errno 
The variation is  
$$
\d{\cal L}_{\rm F}= 
-\left(\dot{\bf A}\hat{S}_{ab}{\bf A}\right)\left(\d\a^{ab}\right)^{\bullet}\,\,.
$$
The variation of the action integral
$$
0=\d I=\int d^4x\d {\cal L}_{\rm F}=-\int d^4 x\left(\dot{\bf A}
\hat{S}_{ab}{\bf A}\right)\left(\d\a^{ab}\right)^{\bullet}\,.
$$
Evaluating this integral by parts and assuming that at infinity the
fields go to zero, the variation is 
$$ \d I=\int
d^4x\left(\d\a^{ab}\right)\left(\dot{\bf A}\hat{S}_{ab} {\bf
A}\right)^{\bullet}\,\,, $$ Using Noether theorem\cite{itzykson}, 
we define the current
$$
\dot{\bf A}\hat{S}_{ab}{\bf A}=\left(\dot{J}_{ab}\right)^0\,\,.
$$
In the world-tube, where we defined the field, the spin tensor
will be
\be
S_{ab}=\int \left(\dot{J}_{ab}\right)^0 d^3x\,\,.
\label{eq:spin-integral}
\ee
Since the spin of the vector field is 1 by definition, we do not
need to integrate (\ref{eq:spin-integral}) over the whole space and can
express spin directly through the orthonormal basis, defined along the
world-tube. Thus, since our spin matrix is
$$
(\hat{S}_{ab})_{cd}=\quart\left(\eta_{ac}\eta_{bd}-
\eta_{ad}\eta_{bc}\right)\,\,,
$$
we obtain the expression for spin current as
$$
\dot{A}^c A^d(\hat{S}_{ab})_{cd}=\half\left(\dot{A}_a A_b-
\dot{A}_b A_a\right)\,\,.
$$
In the world-tube we assume only plane waves and, thus, identify the spin
current with the spin. Thus, our expression for the conserved quantity,
spin of the field, is
$$
S_{ab}=\half \int \left(\dot{A}_a A_b-\dot{A}_b A_a\right)\,d^3x\,\,.
$$
We shall note that the spin tensor $S_{ab}$ is defined now only on a geodesic,
in the same way as velocity and momentum vectors are defined,
which completes our transition from field description to particles.

\vspace*{1pt}\textlineskip

\end{document}